\documentclass[aps,prd,twocolumn,showpacs,superscriptaddress,floatfix,amsmath,amssymb,longbibliography,10pt]{revtex4-1}

\usepackage{color}
\usepackage{dcolumn}
\usepackage{braket}
\usepackage{bm}
\usepackage{bbm}
\usepackage{float}
\usepackage{hyperref}
\usepackage{amsmath}
\usepackage{amssymb}
\usepackage{amsfonts}
\usepackage{graphicx}
\usepackage{comment}
\usepackage{orcidlink}
\usepackage{physics}
\usepackage{multirow}
\usepackage{tgtermes}
\usepackage{ulem}
\usepackage[english]{babel}
\usepackage{txfonts}
\usepackage[altg]{qtxmath}
\usepackage{microtype}
\usepackage{slashed}
\usepackage{diagbox}

\usepackage[utf8]{inputenc}
\usepackage[T1]{fontenc}

\makeatletter
\newcommand*{\rom}[1]{\expandafter\@slowromancap\romannumeral #1@}
\makeatother

\begin{document}

\title{Controlling Schwinger tunneling via engineering of virtual particle phases in vacuum}

\author{D. D. Su \orcidlink{https://orcid.org/0000-0002-9851-7301}}   
\affiliation{Shanghai Normal University, Shanghai 200234, China}
\affiliation{Graduate School, China Academy of Engineering Physics, Beijing 100193, China}
\author{B. F. Shen \orcidlink{https://orcid.org/0000-0003-1021-6991}} 
\affiliation{Shanghai Normal University, Shanghai 200234, China}
\author{Q. Z. Lv \orcidlink{https://orcid.org/0000-0002-7675-0049}}    
\email{qzlv@gscaep.ac.cn} 
\affiliation{Graduate School, China Academy of Engineering Physics, Beijing 100193, China}
\date{\today}

\begin{abstract}
	An investigation into Schwinger pair production mechanisms is presented, demonstrating that vacuum tunneling processes can be effectively controlled through electromagnetic potential modulation while maintaining the strong fields in the interaction region. This challenges the conventional paradigm that attributes exclusive governance of Schwinger processes to localized field intensities. Through comprehensive analysis of particle number, momentum spectra, and spatial distribution of created pairs, we establish that the observed modulation effects originate from electromagnetic potential - induced modifications to the quantum phase structure of virtual particles. This phenomenon reveals a profound connection between Schwinger tunneling dynamics and the geometric phase properties of the quantum vacuum state - a vacuum analogue to the Aharonov-Bohm effect in charged particle systems. This discovery not only advances our understanding of electromagnetic interactions in quantum vacuum but also opens up new experimental opportunities for realizing Schwinger tunneling processes with existing facilities.	
\end{abstract}

\maketitle

\section{Introduction}
The quantum vacuum is far from being an empty void in quantum field theory framework, representing the lowest possible energy level \cite{srednicki2007quantum,weinberg1995quantum}. Unlike the classical concept of a vacuum, it is a dynamic and complex entity filled with fluctuations and virtual particles \cite{bordag2001new,milonni2013quantum,hoi2015probing,PhysRevLett.130.041601}. These fluctuations give rise to a variety of fascinating phenomena \cite{zbMATH03048490,PhysRevA.59.4061,Su_2024}, including the Schwinger tunneling process, a cornerstone prediction of quantum electrodynamics (QED) \cite{fradkin1991quantum,greiner2005quantum}. In strong electromagnetic fields, the vacuum can decay into on-shell electron-positron pairs \cite{RevModPhys.78.591,di2012extremely,AdvancesinQED}, as virtual particles gain sufficient energy to become real. This process, first theorized by Julian Schwinger \cite{PhysRev.82.664}, involves quantum tunneling from negative to positive energy states and remains a vibrant area of research due to its inherently nonperturbative nature and rich theoretical implications \cite{PhysRevLett.111.211603,PhysRevLett.107.171601,GELIS20161,PhysRevLett.116.090406,PhysRevLett.131.021902,PhysRevLett.130.221502,PhysRevLett.133.071803,PhysRevLett.109.253202,PhysRevD.38.348,PhysRevD.38.3593,PhysRevLett.111.060404,hebenstreit2013real,stahl2016schwinger,pla2021pair,jiang2023backreaction,PhysRevLett.130.241501,PhysRevLett.101.130404,PhysRevLett.117.081603,PhysRevD.99.056006,PhysRevResearch.6.023056}. 

Traditionally, the Schwinger effect has been conceptualized as the quantum vacuum functioning as a passive medium that responds to external fields. To initiate Schwinger tunneling, the electric field needs to be the order of $\rm E_{cr} \approx 1.3 \times 10^{18} \thinspace V/m $ \cite{PhysRev.82.664}, which corresponds to a laser intensity $\rm I \approx 4.6 \times 10^{29} \thinspace W/cm^2$ \cite{bulanov2003light,bulanov2010schwinger}. Even with the recent experimental efforts aimed at developing high-powered laser systems, the current world record $\rm I \sim 10^{23} W/cm^2$ \cite{Yoon21} remains several orders of magnitude lower than the required intensity for Schwinger tunneling. Consequently, most of the attention has focused on how additional external electric or magnetic fields might be used to enhance and control the Schwinger mechanism, making it experimentally observable \cite{PhysRevLett.111.060404,PhysRevLett.102.080402,PhysRevLett.103.170403,PhysRevLett.104.220404}. 

All existing approaches primarily focus on modifying the strong fields by increasing field intensity or introducing another field within the interaction region to modify the Schwinger process \cite{nikishov1970pair,karbstein2020probing,PhysRevLett.108.030401,Fillion-Gourdeau_2013,PhysRevLett.110.013002}. The extreme field intensity required makes such modifications experimentally challenging. Introducing additional fields to alter the existing strong field, which often requires precise control and synchronization, is significantly more complex experimentally. Furthermore, high-intensity laser fields inherently exhibit spatial and temporal non-uniformities, complicating the maintenance of a stable and consistent interaction region. These factors collectively make field modification a technically demanding task, highlighting the need for alternative, less invasive novel mechanisms. 

Recent advances in quantum field theory have revealed that electromagnetic potentials can fundamentally influence both the intrinsic properties of the quantum vacuum and the corresponding physical processes \cite{Lv_2016epl,2019Geometric,cetto2022electromagnetic}. Unlike electromagnetic fields that mediate local interactions, potentials provide a more complete description of electromagnetic interactions, as demonstrated by the Aharonov-Bohm (AB) effect in charged particle systems. In AB effect, electromagnetic potentials induce measurable phase shifts in particle wavefunctions even in field-free regions \cite{PhysRev.115.485,PhysRevLett.5.3,vaidman2012role}, establishing potentials as essential physical entities beyond mere mathematical constructs.

This paradigm can be extended to the realm of quantum vacuum physics. The geometric phase structure of virtual particles has been recognized as a crucial factor that governs vacuum dynamics \cite{PhysRevLett.13.508,2006An}. Specifically, the local phase configuration of these virtual particle-antiparticle pairs determines their tunneling probabilities through non-perturbative processes \cite{PhysRevD.53.2190,PhysRevResearch.2.023257,PhysRevD.101.096009,PhysRevD.101.096003}. 

\begin{figure}%[htbp]
	\centering
	\includegraphics[height=7.5cm,width=8.5cm]{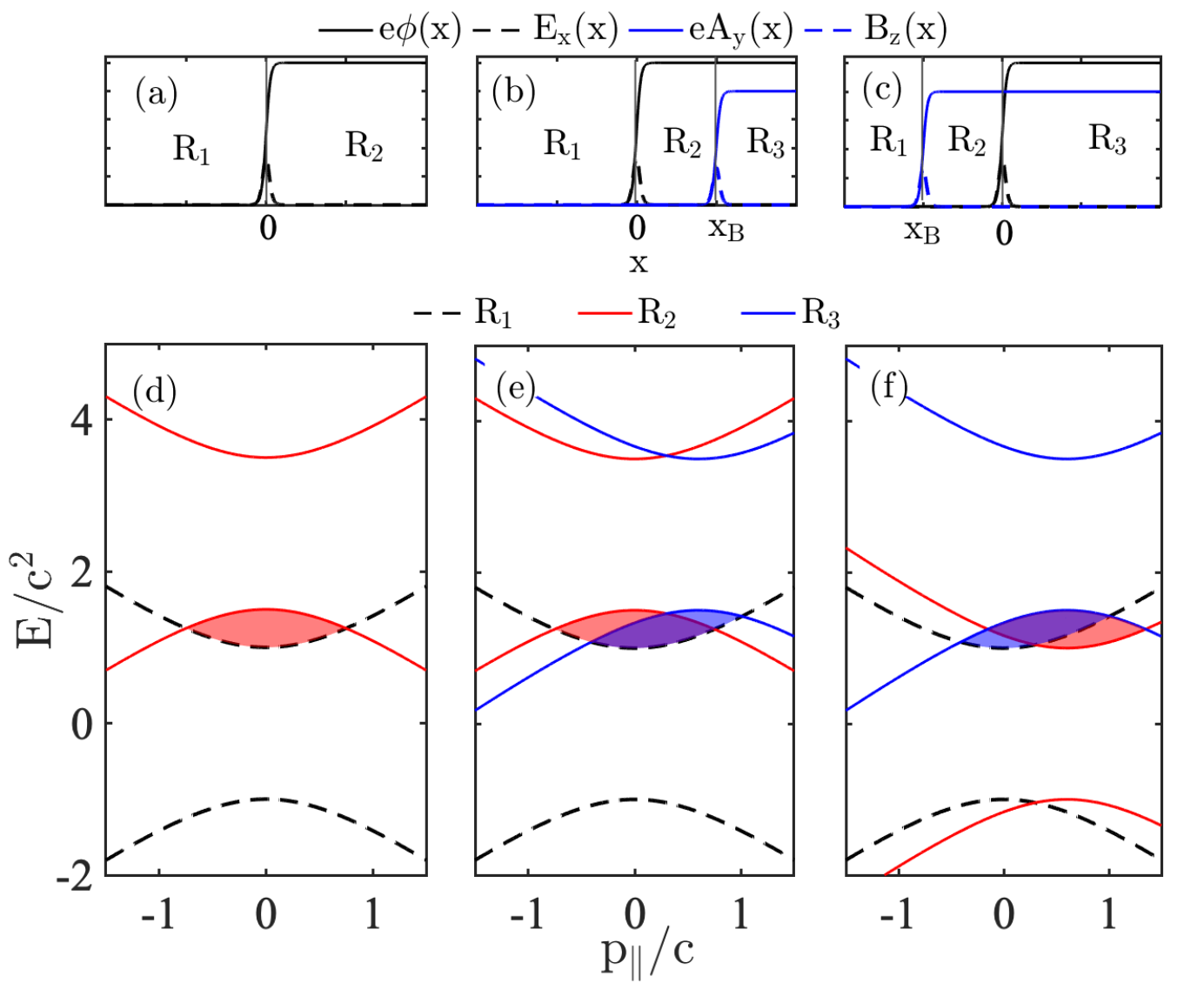}
	\caption {Up row: Sketch of the electromagnetic fields for different configurations. Low row: The corresponding energy spectrum of the Dirac Hamiltonian in different regions. The shadows show the overlap between the positive and negative continuum under the external fields. The parameters for the fields are $\rm e A_0=0.6c^2$, $\rm W_a=0.1 \rm \lambda_c$, $\rm e \phi_0=2.5c^2$, $\rm W_v=0.1 \rm \lambda_c$, and $\rm x_B=\pm L=\pm 24.5\lambda_c = 0.2 a.u.$ for (b) and (c), respectively. Here $\rm \lambda_c = 1/c$ is electron's Compton wave length, which is also the length scale of the pair creation process. The momentum $\rm p_{\parallel}$ means parallel to the vector potential, while the momentum perpendicular to the vector potential is denoted by $\rm p_{\perp}$. } 
	\label{fig:Figure1}
\end{figure}

In this work, our objective is to present an innovative approach for controlling or enhancing the Schwinger tunneling process through engineering electromagnetic potentials. This approach capitalizes on the influence of electromagnetic potentials while keeping the strong electric field within the interaction region intact. The impact of the potential can be mostly observed in the particle number and the momentum spectra of the created particles. Typically, pair creation takes place in a specific energy region, referred to as the Klein region \cite{Hund1941}, where the positive and negative continua overlap [see Fig.~\ref{fig:Figure1}]. By introducing a static vector potential, the Klein region undergoes a shift, which in turn results in a corresponding displacement of the production processes. Moreover, within certain ranges, the momentum distributions display interference patterns [Fig.~\ref{fig:Figure2}], which uncover the subtle role that potentials play in shaping quantum processes. These findings unlock new opportunities for the experimental control of quantum vacuum phenomena and pair productions under extreme conditions.

\section{numerical Results}
Before diving into the quantitative aspects of this new phenomenen, let's first examine the geometry of the field configurations depicted in Fig.~\ref{fig:Figure1}. The first row presents three different setups: Case I [panel (a)] illustrates the traditional Schwinger tunneling configuration with a single electric field at the center, where the field strength is sufficient to cause an overlap between the positive and negative continua; Case II [panel (b)] introduces an additional vector potential, positioned far from the pair creation region around $\rm x=0$; Case III [panel (c)] also includes a vector potential, but this time it overlaps with the interaction zone. It is worth to point out that in Case II and Case III the magnetic fields induced by the vector potentials are far from the interaction region. The second row of Fig.~\ref{fig:Figure1} shows the boundaries of the positive ($\rm E_+$) and negative ($\rm E_-$) continua, derived from the eigenvalues of the static Dirac equation $\rm \hat{H}\psi=E\psi$. Here, $\psi$ represents the wave function of the virtual particles and the Dirac Hamiltonian reads 
\begin{equation}
	\label{eq:eq1}
	\hat{\rm H}=c \boldsymbol{\alpha}  \cdot (\hat{\boldsymbol{p}}-e \boldsymbol{A}/c)+\beta m c^2 +e\phi .
\end{equation}
Here, $\boldsymbol{\alpha}=(\alpha_1,\alpha_2,\alpha_3)$ and $\beta$ denote the set of four $4 \times 4$ Dirac matrices, while $\boldsymbol{A}$ and $\phi$ represent the vector and scalar potentials of the electrodynamics fields, respectively. $m$ and $e$ are the electron mass and charge. Henceforth, we adopt atomic units (a.u.) with $\hbar=m=|e|=1$, unless stated otherwise. The electric and magnetic fields were modeled using the scalar potential $\rm \phi(x)=\phi_0\{1+\tanh[x/W_v]\}/2$ and the vector potential $\rm A_y(x)=A_0\{1+\tanh[(x+x_B)/W_a)]\}/2$ along the y-axis. This configuration generates an electric field along the x-axis and a magnetic field along the z-axis. 

In the context of quantum field theory, the vacuum state is filled with virtual particles. The electromagnetic potential can interact with these virtual particles, modifying their properties. As a result, the properties of the vacuum state are also affected. For example, the introduction of the vector potential can modulate the phase of the wave function $\psi$. This modulation alters the Klein region and the corresponding Schwinger processes, seen in the Appendix. This phenomenon exhibits an analogy to the traditional AB effect. In both situations, it is the electromagnetic potential rather than the electromagnetic field that influences the physical processes.  

In Case I [Fig.~\ref{fig:Figure1}(d)], the energy levels undergo a vertical shift for $\rm x>0$. The Klein region is therefore defined by the intersection of the black dashed line and the red solid line. Moving on to Case II [Fig.~\ref{fig:Figure1}(e)], the presence of the vector potential induces a shift in the spectrum with respect to $\rm p_\parallel$ in the $\rm R_3$ region. This shift causes the Klein region, represented by the overlap of the red and blue shaded areas, to lose its symmetry about $\rm p_\parallel$. Additionally, the purely red shaded small region represents the overlap between the positive energy states for $\rm R_1$ region and the negative energy states for $\rm R_2$ region. Since the negative energy states in the $\rm R_2$ region exhibit certain resonant energies, they provide distinct channels for pair production outside the true Klein region. The purely blue shaded area corresponds to the overlap between $\rm R_1$ and $\rm R_3$ regions, a situation that cannot occur physically. Case III bears resemblance to Case II, with the only difference being the position of the vector potential [Fig.~\ref{fig:Figure1}(c)]. Nevertheless, the Klein region is altered, as evident in Fig.~\ref{fig:Figure1}(f). The resonant energy states in $\rm R_2$ once again offer individual channels for pair creation, similar to Case II. All of these resonant creation channels are reflected in the production process.

In order to simulated the pair production, we employ the computational quantum field theory (CQFT) approach \cite{greiner2005quantum,PhysRevD.96.056017}. This approach enables us to compute the particle number as well as the spatial and momentum distribution of the created electron-positron pairs. Moreover, unlike traditional QED calculations, CQFT also allows us to track the time evolution of the process, providing an additional degree of freedom to study the Schwinger tunneling mechanism. Based on the CQFT approach, the quantum field operator can be expanded using time-dependent or independent basis as $\Psi(\rm t)= \rm \sum_p b(p,t)\ket{p}+ \sum_n d^{\dagger} (n,t)\ket{n} = \sum_p b(p)\ket{p(t)}+ \sum_n d^{\dagger}(n)\ket{n(t)}$, where $\rm \ket{p}$ and $\rm \ket{n}$ represent the positive and negative eigenstates of the field-free Dirac Hamiltonian, respectively, while $\rm \ket{p(t)}$ and $\rm\ket{n(t)}$ are solutions to the Dirac equation in the presence of background fields. The electronic creation and annihilation operators [$\rm b^{\dagger}(p)$ and $\rm b (p)$] and positronic operators [$\rm d^{\dagger}(p)$ and $\rm d(p)$] satisfy the anti-commutation relations: $\rm \{b(p),b^\dagger (p')\} = \{ d(p),d^{\dagger} (p') \} =\delta_{p,p'}$, with all other commutators vanishing. Thus, the electron momentum distribution (EMD) can be determined by evaluating the expectation value of the creation and annihilation operators as:
\begin{equation}
	\label{eq:eq2}
	\rm \rho (p,t)=\bra{\bra{vac}} b^\dagger(p,t)b(p,t)\ket{\ket{vac}}=\sum_n |G_{p,n}(t)|^2,
\end{equation}
where the last equality follows from the generalized Bogoliubov transformation, with the matrix element $\rm G_{p,n}(t)=\braket{p}{n(t)}=\bra{p} U(t) \ket{n}$ involving the time evolution operator $\rm U(t)$ associated with the Dirac Hamiltonian $\rm \hat{H}$ in Eq.~\eqref{eq:eq1}. The quantum vacuum state is denoted by the double ket notation $\rm \ket{\ket{vac}}$. The total particle number is then obtained as $\rm N(t)=\sum_{p,n} |G_{p,n}(t)|^2$ and the energy spectrum as $\rm \rho (E,t)=\rho (p,t)dp/dE$ with $\rm E =(\rm c^4+p^2c^2)^{1/2}$. 

The electromagnetic fields depicted in Fig.~\ref{fig:Figure1} are all localized along the x-axis. This localization ensures the conservation of momentum $\rm p_y$ and $\rm p_z$ throughout the entire process. Additionally, considering that the electric field is aligned along the x-axis and the magnetic field along the z-axis, there is no dynamics along the z-axis. Without loss of generality, we can set $\rm p_z=0$. This simplification effectively reduces our ab initial simulation from a three-dimensional system to a series of one-dimensional systems, rendering it numerically feasible.

\begin{figure}%[htbp]
	\centering
	\includegraphics[height=4.0cm,width=8.5cm]{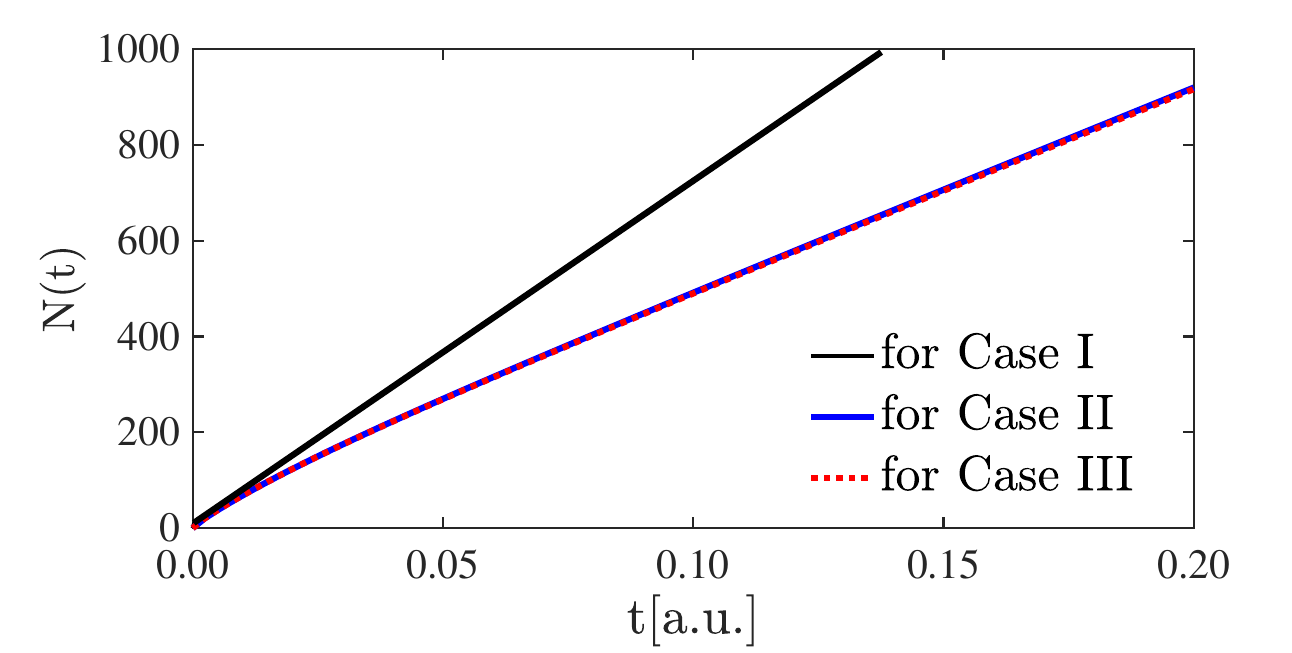}
   	\caption {The particle number as a function of time for the created electrons in all three cases. All the parameters are the same as in Fig.~\ref{fig:Figure1}. } 	
	\label{fig:Figure1-2}
\end{figure}

Let us now discuss the numerical results. As depicted in Fig.~\ref{fig:Figure1}, the Klein regions in Case II and Case III are smaller than that in Case I. This indicates that the pair creation rate in Case I is higher than in Case II and Case III. The temporal evolution of the particle number, presented in Fig.~\ref{fig:Figure1-2}, corroborates this trend. Specifically, the long-time pair creation rate in Case I is $\Gamma_{\rm I}=7251$, approximately $40\%$ greater than that in Case II ($\Gamma_{\rm II}=4260$) and Case III ($\Gamma_{\rm III}=4244$). The creation rates in Case II and Case III are comparable due to the similar sizes of their Klein regions [Fig.~\ref{fig:Figure1}(b) and (c)]. The creation rate as a function of $\rm p_{\parallel}$ also showes that the Klein regions for Case II and Case III are different, yet have similar total sizes. This disparity can be exploited in experiments to suppress pair creation under certain conditions, thereby conserving laser energy. Additionally, the creation rates in Case II and Case III are not constant over short time and long time scales, as evident from the blue and red curves in Fig.~\ref{fig:Figure1-2}. This behavior is attributed to the resonant pair creation channels (in region $\rm R_2$) for short time scale, as previously discussed. 

\begin{figure}%[htbp]
	\centering
	\includegraphics[height=7.0cm,width=8.5cm]{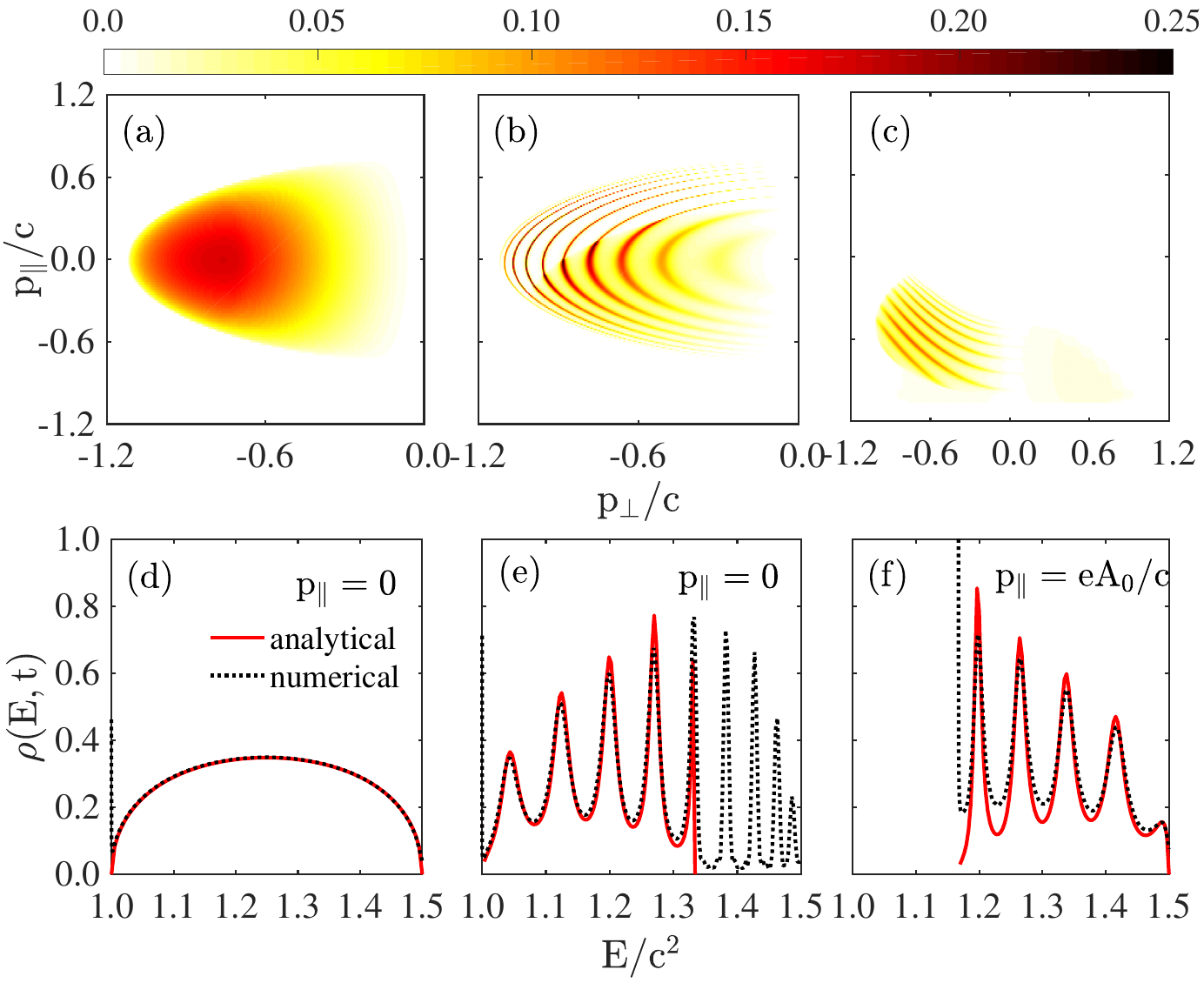}
   	\caption {Top row: The normalized EMD for the three cases as shown in Fig.~\ref{fig:Figure1} with an interaction time of $\rm t=0.036 a.u.$. Bottom row: The energy spectra of the created electrons with the most probable $\rm p_\parallel$. The dotted curves represent the numerical results obtained from the EMD cut at the most probable $\rm p_\parallel$, while the solid lines are the analytical predictions based on Eq.~(\ref{eq:eq3-1}-\ref{eq:eq3-3}). All other parameters are the same as in Fig.~\ref{fig:Figure1}. } 	
	\label{fig:Figure2}
\end{figure}

To conduct a more in-depth analysis of the tunneling processes, we present the two-dimensional EMDs for the three cases in Fig.~\ref{fig:Figure2}. Panel (a), corresponding to Case I, illustrates that the created electrons exhibit symmetric parallel momentum $\rm p_\parallel$, which is consistent with the symmetric Klein region in Fig.~\ref{fig:Figure1}(d). Due to the influence of the electric field direction, electrons move left after creation. As a result, the perpendicular momentum $\rm p_\perp$ shifts leftward. The range of $\rm p_\perp$ can be estimated using the formula $\rm p_\perp=(\rm E^2-c^4-p_\parallel^2c^2)^{1/2}$, where $\rm E$ represents the energy within the Klein region. By substituting the values obtaining from Fig.~\ref{fig:Figure1}(d) into this formula, we obtain a range of $\rm -1.118c < p_\perp<0$. This range matches the range presented in Fig.~\ref{fig:Figure2}(a).

\begin{figure*}[!htbp]
	\centering
	\includegraphics[height=9.0cm,width=\textwidth]{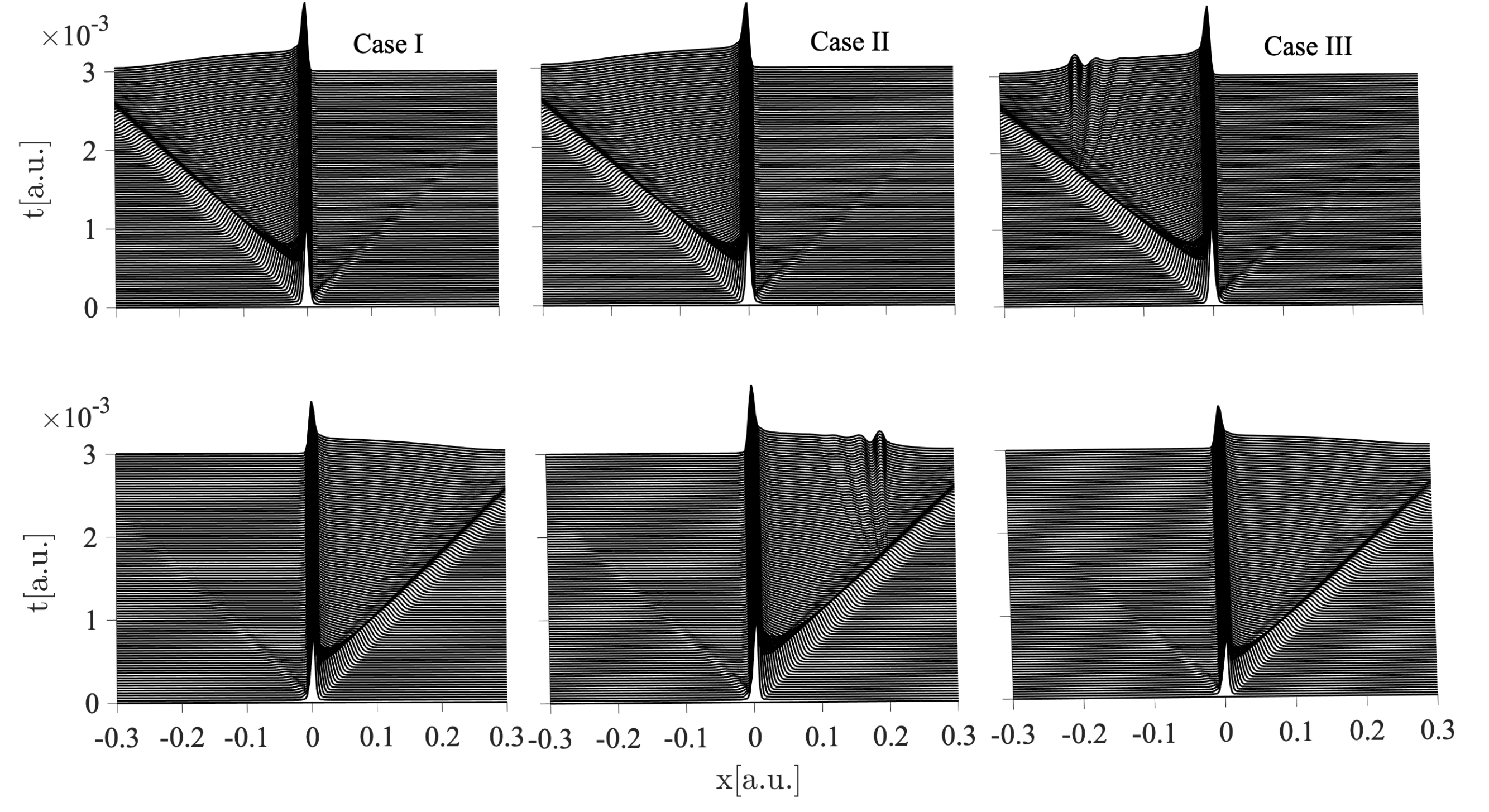}
   	\caption {The time evolution of the spatial distribution $\rm \rho(x,t)$ for the created electrons (first row) and positrons (second row) in all three cases. The other parameters are the same as in Fig.~\ref{fig:Figure1}. } 	
	\label{fig:Figure2-3}
\end{figure*}

Fig.~\ref{fig:Figure2}(b) for Case II demonstrates the influence of the vector potential. Notably, this influence persists even though the magnetic field does not overlap with the interaction region around $\rm x=0$. The EMD exhibits a range roughly similar to that in Fig.~\ref{fig:Figure2}(a). However, it features distinct interference patterns. These patterns stem from the superposition of left and right going particles in region $\rm R_2$, as depicted in Fig.~\ref{fig:Figure1}(b). We will provide a more detailed explanation of this interference phenomenon within the analytical model below.

Furthermore, two distinct regimes are observable in the corresponding energy spectrum shown in Fig.~\ref{fig:Figure2}(e). The energy range from $\rm c^2$ to $\rm 1.33c^2$ corresponds to the Klein region, as illustrated in Fig.~\ref{fig:Figure1}(e). In this range, the creation mechanisms are driven by both the tunneling process and the interference enhancement. Intuitively, this can be interpreted as negative energy states originating from $\rm x=+\infty$ tunneling through the region between the electric and magnetic fields and transitioning to positive energy states at $\rm x=-\infty$. When the energy of the incoming states coincides with the resonant energy in region $\rm R_2$, the tunneling process is enhanced. The second regime, spanning from $\rm 1.33c^2$ to $\rm 1.5c^2$, aligns with the pure red shaded region in Fig.~\ref{fig:Figure1}(e). In this regime, individual resonance channels remain present, even though tunneling is no longer feasible.

Even more astonishing is the EMD presented in Fig.~\ref{fig:Figure2}(c). It exhibits marked differences when compared to both Figs.~\ref{fig:Figure2}(a) and (b). Conventionally, based on the well established correlations between created electrons and positrons\cite{PhysRevD.95.036006,PhysRevLett.121.183606}, relocating the magnetic field from $\rm x=x_B=L$ to $\rm x=x_B=-L$ should not impact the EMDs. However, the Klein region in Fig.~\ref{fig:Figure1}(f) has shifted entirely relative to that in Fig.~\ref{fig:Figure1}(e). This shift can be attributed to the overlap between the vector potential $\rm A_y(x)$ and the interaction region, as illustrated in Fig.~\ref{fig:Figure1}(c). The contrast between the EMDs in Fig.~\ref{fig:Figure2}(b) and (c) further validates that electromagnetic potentials are more fundamental than electromagnetic fields. Beyond this shift, another notable difference is that some electrons are moving in the positive x-direction. This occurs because the magnetic field around $\rm x_B=-L$ alters the propagation direction of some electrons after they scattering in the magnetic field region. This phenomenon is also evident in the time evolution of the spatial distribution shown in Fig.~\ref{fig:Figure2-3}. Moreover, the EMD in panel (c) displays an interference pattern superimposed on the tunneling process.

\section{Analytical model}
The interference in Case II and III originates from the same mechanism. To understand this in more details, we examine the energy spectrum of the created electrons with the most probable $\rm p_\parallel$ in the EMD and compare them with an analytical model. This model relates the energy spectrum of the electrons to the quantum mechanical transmission coefficient of an incoming wave packet \cite{Hund1941,PhysRevA.86.013422}. Assuming a $\delta$-function distribution in along x-axis for both electric and magnetic fields in Fig.~\ref{fig:Figure1}, the transmission coefficient can be obtained from the stationary energy eigenstates as shown in the Appendix. Then the energy distribution for all three cases can be written as $\rm \rho_{I,II,III} (E,t)= 2t/\pi T_{I,II,III}(E)$, where the transmission coefficients for three cases, respectively, look like  
\begin{subequations}
	\begin{align}
		& \rm T_I(E_i) = \frac{\rm c^2 p_{i,\perp} p_{f,\perp} (E_i+c^2)(E_f+c^2)}{\rm (p_{i,\perp} +p_{f,\perp})^2 p_{i,\parallel}^2+[(E_i+c^2)(E_f+c^2)+p_{i,\perp} p_{f,\perp}-p_{i,\parallel}^2]^2} \label{eq:eq3-1}  \\
		%& \rm T_I(E_i) = \frac{\rm 2c^2 p_{i,\perp} p_{f,\perp}}{\rm E_i E_f + c^4 + c^2 p_{i,\perp} p_{f,\perp} + c^2 p_{i,\parallel} p_{f,\parallel}} \label{eq:eq3-1}  \\
		& \rm T_{II}(E_i) = \frac{\rm c^2 p_{i,\perp} p_{2,\perp}^2 p_{f,\perp} (E_i+c^2)(E_f+c^2)}{\rm [c_a \sin(\eta)+c_b \cos(\eta)]^2+[c_c \sin(\eta)+c_d \cos(\eta)]^2} \label{eq:eq3-2}  \\
		& \rm T_{III}(E_i) = \frac{\rm c^2 p_{i,\perp} p_{2,\perp}^2 p_{f,\perp} (E_i+c^2)(E_f+c^2)}{\rm [c_a' \sin(\eta)+c_b' \cos(\eta)]^2+[c_c' \sin(\eta)-c_d' \cos(\eta)]^2} \label{eq:eq3-3}
	\end{align}
\end{subequations}
where $\rm \eta=p_{2,\perp} L$. The quantities $\rm E_i$, $\rm p_{i,\perp}$, and $\rm p_{i,\parallel}$ pertain to the incoming particles in region $\rm R_1$ of Fig.~\ref{fig:Figure1}. In region $\rm R_2$, the corresponding quantities are $\rm E_2$, $\rm p_{2,\perp}$, and $\rm p_{2,\parallel}$. In region $\rm R_3$, they are $\rm E_f$, $\rm p_{f,\perp}$, and $\rm p_{f,\parallel}$ for the outgoing particles. The relations between them are provided in the Appendix. From Eq.~\eqref{eq:eq3-2} and Eq.~\eqref{eq:eq3-3}, it becomes evident that the transmission coefficients for Case II and Case III exhibit a similar mathematical form. The only difference lies in the coefficients $A,B,C$ and $D$ in the denominator.

For Case I with only electric field, Eq.~\eqref{eq:eq3-1} shows that the energy spectrum is a semi-circle. For Case II and III, when the vector potential is present, the transmission coefficients have some oscillating terms proportional to $\sin(\eta)$ and $\cos(\eta)$ emerging in the denominator in Eq.~\eqref{eq:eq3-2} and Eq.~\eqref{eq:eq3-3}. These terms is caused by the interference between the left and right going waves in region $\rm R_2$, see the derivation in the Appendix. In Fig.~\ref{fig:Figure2}(d-f), the agreement between the exact numerical spectra and the analytical predictions is excellent. The only difference is that in Fig.~\ref{fig:Figure2}(e) the peaks for high energy in numerical spectrum are not reproduced by the analytical model. This associated with the fact that the analytical expression describes the tunneling process from $ \rm x=-\infty $ to $\rm x=+\infty$ while the numerical spectrum also includes the resonant channels in region $\rm R_2$ as mention before. For Case III, however, there is no overlap between the positive and negative energy states in the region of $\rm R_2$ and thus no individual pair creation channels outside of the Klein region. Moreover, by integrating over the whole momentum space for the transmission coefficient, we can obtain the total creation rate for all cases as $\Gamma_{\rm I,A}=7178$, $\Gamma_{\rm II,A}=4158$, and $\Gamma_{\rm III,A}=4135$. They all agree with the numerical results shown in Fig.~\ref{fig:Figure1-2} within $2.65\%$. This provides the validation that the EMDs calculated above can accurately represent the long-term asymptotic behavior of the pair creation process. 

\begin{figure}%[htbp]
	\centering
	\includegraphics[height=7.5cm,width=8.5cm]{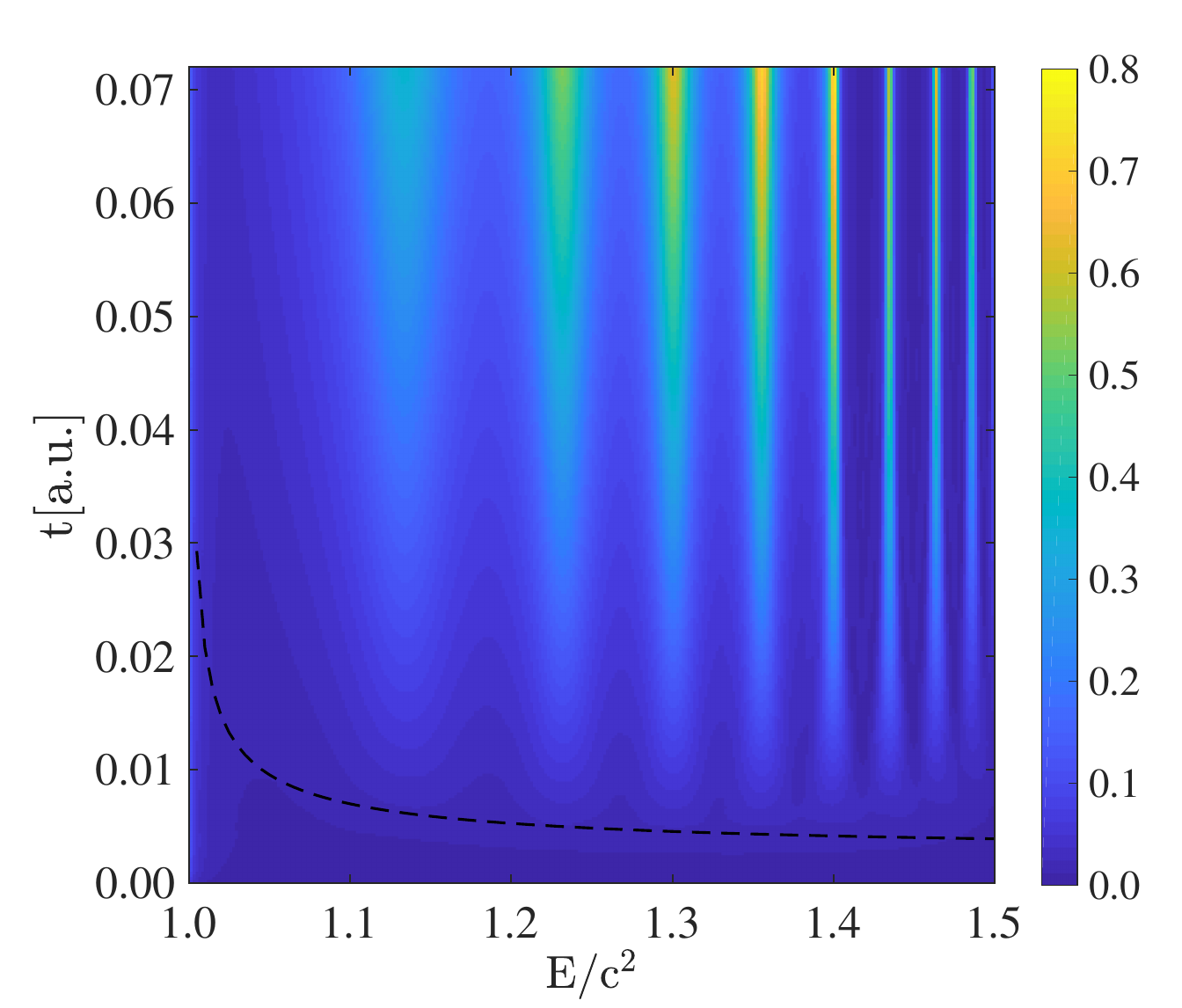}
   	\caption {The energy spectrum of the created electrons for the most probable longitudinal momentum $\rm p_\parallel=0$ as a function of time for Case II. The black dashed curve indicates the time when the corresponding positrons turn back at the vector potential ($\rm x=L$) and arrive at $\rm x=0$. The other parameters are the same as in Fig.~\ref{fig:Figure1}. } 	
	\label{fig:Figure3}
\end{figure}

Insight into the impact of the phase of virtual particles on the Schwinger process can be significantly enhanced by examining the time evolution of the spatial and energy distribution of created pairs. This offers a novel degree of freedom for studying the pair production process, which is possible only within the framework CQFT approach, with $\rho({\rm x,t})=\sum_{\rm p} \bra{\bra{{\rm vac}}} \Psi_{\pm,\rm p}^\dagger({\rm x,t})\Psi_{\pm,\rm p}(\rm x,t)\ket{\ket{\rm vac}}$ \cite{PhysRevD.96.056017}. Here, $\Psi_{\pm,\rm p}^\dagger(\rm x,t)$ represents the electron(+)/positron(-) part of the field operator. Fig.~\ref{fig:Figure2-3} depicts the spatial distributions of created electrons and positrons for all cases. In Case I, electrons and positrons move symmetrically to the left and right immediately after creation. In Case II, since the vector potential is positioned to the right of the strong electric field, electrons do not interact with the vector potential throughout the process. Conversely, positrons are scattered by the corresponding magnetic field post-creation. The superposition of positrons moving to the left and right in this region ($\rm 0<x<L$) will cause interference pattern in the positron momentum distribution. Due to the correlation between electrons and positrons, the EMD also exhibits an interference pattern [Fig.~\ref{fig:Figure2}(b)]. As shown in the second panel of the second row here, positrons reach the magnetic field region at approximately $\rm t_{inf}=2\times10^{-3}$ a.u.. This marks the time scale at which the interference pattern in the EMD begins to emerge. Case III follows a similar process, with the roles of electrons and positrons reversed. Overall, these spatial distributions confirm that particles are created exclusively within the strong electric field region. The vector potential influences the Schwinger tunneling process solely through the delocalized phase modulation of virtual particles in the vacuum. This finding is consistent with the analytical model, where the phase of the negative continuum wave function is altered.

Figure.~\ref{fig:Figure3} also presents the time evolution of the energy spectrum for the created electrons in Case II, with the most probable $\rm p_\parallel$. Analysis of this figure reveals two distinct time domains in the processes. Initially, electron production primarily occurs via the tunneling process within the Klein region. Post-creation, as positrons are scattered by the magnetic field, interference peaks also begin to emerge in the electron's energy spectrum. The black dashed curve in the figure marks the onset of interference, comparable to $\rm t_{inf}$ in Fig.~\ref{fig:Figure2-3}. As time progresses, the interference phenomenon becomes increasingly prominent.

From these observations, we can infer that the shift of the Klein region is induced by changes in the phase of virtual particles in the vacuum and is instantaneous. In contrast, the interference phenomenon represents a dynamical process. It arises from the superposition of left and right moving particles and can only develop a certain time after the particle being scattered by the magnetic field.

\section{Conclusion}
In summary, this study demonstrates that external electromagnetic potentials can modulate the local geometric phase of virtual particles in the quantum vacuum. The interference pattern observed in the momentum distribution effectively enhances pair creation within certain momentum intervals. This property can be exploited to detect the Schwinger process. Experimentally, our proposed setup is, in principle, realizable in heavy ion collision experiments. During head-on or grazing collisions between two heavy ions, the electromagnetic vector potential interacts (or does not interact) with the Coulomb potential. This interaction generates a field configuration akin to that depicted in Fig.~\ref{fig:Figure1}. This research underscores the active role of the vacuum's internal structure. By elucidating the relationship between electromagnetic potentials and virtual particles, this work opens a new avenue for fine-tuning Schwinger tunneling processes and lays a solid foundation for future investigations into quantum vacuum phenomena. 

\section*{ACKNOWLEDGMENTS}
DDS would like to thank Prof. Lv for the nice hospitality during her visiting at GSCAEP in Beijing. This work was supported by National Natural Science Foundation of China (NSFC) (Grant No. 12304341, No. 11935008, and No. 12475251). Simulations were performed on the $\pi$ supercomputer at Shanghai Jiao Tong University.

\section*{Appendix I: an analysis of a quantum mechanical scattering process}
\label{appendix1}
In the main text, we developed an analytical model to describe a quantum mechanical scattering process in two-dimensional space, which can be related to the Schwinger process through Hund formula\cite{Hund1941,PhysRevA.86.013422}. Here, we provide a completely analytical derivation of the quantum mechanical two-dimensional scattering process for a particle on various configurations of electric and magnetic fields and the transmission coefficient is obtained at the end.

The field configurations are similar as sketched in Fig.~1 of the main text. However, in the analyical analysis we choose the electric and magnetic field to be a delta distribution in space but keep the magnitude of the corresponding scalar and vector potentials unchanged. Therefore, the width of the fields $\rm W_v=W_a=0$, which simplifies the derivation of the transmission coefficient. 

The scattering states for the Dirac equation can be written as:
\begin{equation}
	\label{eq:psi_plus}
	\begin{split}
		\rm \psi_{+,p,+1/2}(x)=\sqrt{\rm \frac{E_p+mc^2}{2E_p}}
        \begin{pmatrix}
        	1 \\0\\ \rm \frac{cp_z}{E_p+mc^2} \\ \rm \frac{cp_\perp+icp_\parallel}{E_p+mc^2}
        \end{pmatrix}
        \rm exp[i p \cdot x] \,, \\
		\rm \psi_{+,p,-1/2}(x)=\sqrt{\rm \frac{E_p+mc^2}{2E_p}}
		\begin{pmatrix}
			0\\1\\\rm \frac{cp_\perp-icp_\parallel}{E_p+mc^2} \\ \rm \frac{-cp_z}{E_p+mc^2}
		\end{pmatrix}
		\rm exp[i p \cdot x] \,,
	\end{split}
\end{equation}

\begin{equation}
	\label{eq:psi_minus}
	\begin{split}
		\rm \psi_{-,p,+1/2}(\mathbf{x})=\sqrt{\rm \frac{E_p+mc^2}{2E_p}}
        \begin{pmatrix}
        	\rm \frac{cp_z}{E_p+mc^2} \\\rm \frac{cp_\perp+icp_\parallel}{E_p+mc^2}\\1 \\0
        \end{pmatrix}
        \rm exp[-i p \cdot x] \,, \\
		\rm \psi_{-,p,-1/2}(x)=\sqrt{\rm \frac{E_p+mc^2}{2E_p}}
        \begin{pmatrix}
        	\rm \frac{cp_\perp-icp_\parallel}{E_p+mc^2}\\\rm \frac{-cp_z}{E_p+mc^2}\\0 \\1
        \end{pmatrix}
        \rm exp[-i p \cdot x] \,.
	\end{split}
\end{equation}
Here Eq.~\eqref{eq:psi_plus} denotes the positive energy states with spin of $\rm s=\pm 1/2$, while Eq.~\eqref{eq:psi_minus} is the corresponding negative energy states. Because the field configurations studied in the main text cannot coupled different spin, we select $\rm \psi_{+,p,+1/2}(x)$ and $\rm \psi_{-,p,+1/2}(x)$ for the subsequent derivations with the polarization of spin-up along z-axis.

\subsection{Quantum transmission coefficient for case I } 
	\label{qtc1}

In this section, we derive the transmission coefficient $\rm T(E)$ for case I in Fig.~1(a) in the main text, in which there is a step scalar potential around $\rm x=0$. Therefore, in the region $\rm x<0$, the wave function includes both incident and reflected waves, while for $\rm x > 0$, it contains only the transmitted waves,
\begin{equation} 
	\label{eq:psi_i}
	\rm \psi(x)=
		\begin{cases}
			\rm \psi_{+,p_{i,\perp},+1/2}(x)+ r \psi_{+,-p_{i,\perp},+1/2}(x), & x<0 \\
			\rm t \psi_{-,p_{f,\perp},+1/2}(x), & x >0.
	\end{cases}
\end{equation}
The continuity condition for the wave function at $\rm x=0$ yields equations for the coefficient $\rm r$ and $\rm t$:
\begin{eqnarray} 
	\label{eq:con_eq1}
    \begin{aligned} 
       & \rm 1 + r = t \sqrt{\rm \frac{E_i (E_f + c^2)}{E_f (E_i + c^2)}} \frac{cp_{f,\perp}}{E_f+c^2} \,, \\
       & \rm \frac{cp_{i,\perp}-icp_{i,\parallel}}{E_i+c^2} - r \frac{cp_{i,\perp}-icp_{i,\parallel}}{E_i+c^2} = t \sqrt{\rm \frac{E_i (E_f + c^2)}{E_f (E_i + c^2)}}  \,,
    \end{aligned}
\end{eqnarray}
where $\rm p_{i,\perp}$, $\rm p_{i,\parallel}$ and $\rm E_i$ is the momentum and energy for the incoming particles in region $\rm R_1$ and $\rm p_{f,\perp}$, $\rm p_{f,\parallel}$ and $\rm E_f$ for the outgoing particles in region $\rm R_2$ in Fig.~1(a) in main text. Solving Eqs.~\eqref{eq:con_eq1}, the transmission coefficient $\rm T(E)$ is found to have the following form
\begin{eqnarray}
	\label{eq:tran_coef1}
	\begin{aligned}
		\rm T_I(E_i) & = \rm [4c^2 p_{i,\perp} p_{f,\perp} (E_i+c^2)(E_f+c^2)]/\{(p_{i,\perp}+p_{f,\perp})^2   \\ 
		&\rm p_{i,\parallel}^2 + [(E_i+c^2)(E_f+c^2)+p_{i,\perp} p_{f,\perp}-p_{i,\parallel}^2]^2\},
	\end{aligned}
\end{eqnarray}
with $\rm E_f=e\phi_0-E_i$ and $\rm c p_{f,\perp}=(E_f^2-c^2 p_{f,\parallel}^2-c^4)^{1/2}$. The longitudinal momentum $\rm p_{i,\parallel}$ and $\rm p_{f,\parallel}$ are the same in region $\rm R_1$ and $\rm R_2$.

\subsection{Quantum transmission coefficient for case II } 
	\label{qtc2}

The derivation in this section corresponds to case II in the main text. In this scenario, the scalar and vector potentials are modeled by $\rm \phi(x) = \phi_0 [1+\theta(x)]$ and $\rm A_y(x) = A_0 [1+\theta(x+x_B)]$ with $\rm x_B=L$, respectively. These spatially separated scalar and vector potentials divide the space into three regions, i, ii, and iii as shown in Fig.~1(b) of the main text. As the multi-scattering process between $\rm x=0$ and $\rm x=L$ is negligible, we consider only the first-order reflection and transmission processes, the wave functions in these three regions can be expressed as:
\begin{equation} 
	\label{eq:psi_ii}
	\rm \psi(x)=
	\begin{cases}
		\rm \psi_{+,p_{i,\perp},+1/2}(x)+ r \psi_{+,-p_{i,\perp},+1/2}(x), & \rm x<0 \\
		\rm c_1 \psi_{-,p_{2,\perp},+1/2}(x)+ c_2 \psi_{-,-p_{2,\perp},+1/2}(x), & \rm 0<x<L \\
		\rm  t \psi_{-,p_{f,\perp},+1/2}(x), & \rm x>L.
	\end{cases}
\end{equation}
In the region $\rm x<0$, the wave function consists of the incident and the reflected wave by the scalar potential. In the region $\rm 0<x<L$, the wave function is a superposition of the transmitted and the reflected wave by the scalar and vector potential. Finally, in the region $\rm x>L$, the wave function consists solely of the transmitted wave. Although the space is divided into three parts by the scalar potential $\rm \phi(x)$ at $\rm x = 0$ and the vector potential $\rm A_y(x)$ at $\rm x_B = L$, the wave function must be continuous at these two points and the following equations must be fulfilled:
\begin{eqnarray}
	\label{eq:con_eq2}
	\begin{aligned} 
        & \rm \sqrt{\rm \frac{E_2 (E_i + c^2)}{E_i (E_2 + c^2)}} \frac{1 + r}{c} = \frac{ c_1 (p_{2,\perp} - i p_{2,\parallel})}{E_2 + c^2} - \frac{c_2 (p_{2,\perp} + i p_{2,\parallel})}{E_2 + c^2} \,, \\
        & \rm \frac{ c_1 (p_{2,\perp} - i p_{2,\parallel})}{E_2 + c^2} e^{-i p_{2,\perp} L} - \frac{c_2 (p_{2,\perp} + i p_{2,\parallel})}{E_2 + c^2}  e^{i p_{2,\perp} L} \\ 
		& \rm =t \sqrt{\rm \frac{E_2 (E_f + c^2)}{E_f (E_2 + c^2)}} \frac{p_{f,\perp} - i p_{f,\parallel}}{E_f + c^2} \,, \\ 
        & \rm \sqrt{\rm \frac{E_2 (E_i + c^2)}{E_i (E_2 + m c^2)}} \big[\frac{p_{i,\perp} + i p_{i,\parallel}}{E_i + c^2} - r \frac{p_{i,\perp} - i p_{i,\parallel}}{E_i + c^2} \big] = \frac{c_1  + c_2}{c} \,, \\
        & \rm c_1 e^{-i p_{2,\perp} L} + c_2 e^{i p_{2,\perp} L}  = t \sqrt{\rm \frac{E_2 (E_f + c^2)}{E_f (E_2 + c^2)}} e^{-i p_{f,\perp} L},
	\end{aligned} 
\end{eqnarray}
where the momentum and energy in different regions are like $\rm cp_{i,\perp}=(E^2-c^2p_{i,\parallel}^2-c^4)^{1/2}$ for region $\rm R_1$, $\rm E_2=e\phi_0-E$, $\rm c p_{2,\perp}=(E_2^2-c^2p_{2,\parallel}^2-c^4)^{1/2}$ for region $\rm R_2$ with $\rm p_{2,\parallel}=p_{i,\parallel}$, and $\rm E_f=E_2$, $\rm c p_{f,\perp}=(E_f^2-c^2p_{f,\parallel}^2-c^4)^{1/2}$ for region $\rm R_3$ with $\rm p_{f,\parallel}=p_{i,\parallel}-eA_0$. Then, the transmission coefficient is
\begin{eqnarray}
	\label{eq:tran_coef2}
	\begin{aligned}
		\rm T_{II}(E_i) & = \rm [4c^2 p_{i,\perp} p_{2,\perp}^2 p_{f,\perp} (E_i+c^2)(E_f+c^2)]/  \\ 
		&\{\rm [c_a \sin(\eta)+c_b \cos(\eta)]^2+[c_c \sin(\eta)+c_d \cos(\eta)]^2\},
	\end{aligned}
\end{eqnarray}
with the coefficients as  $\rm c_a=p_\perp(E_f+mc^2)(E_2-mc^2)+p_{f,\perp}(E_i+c^2)(E_2+c^2)$, $\rm c_b=p_{2,\perp} c^2 (p_{i,\parallel} p_{f,\perp}+p_{i,\perp} p_{f,\parallel})$, $\rm c_c =p_{i,\parallel} (E_f+c^2)(E_2-c^2)-p_{2,\parallel}(E_i+c^2)(E_f+c^2)+p_{f,\parallel}(E_i+c^2)(E_2+c^2)$ and $\rm c_d = p_{2,\perp}[(E_i+c^2)(E_f+c^2)+c^2 (p_{i,\perp} p_{f,\perp}-p_{i,\parallel} p_{f,\parallel})]$.

\begin{figure}%[htbp]
	\centering
	\includegraphics[height=7.5cm,width=8.5cm]{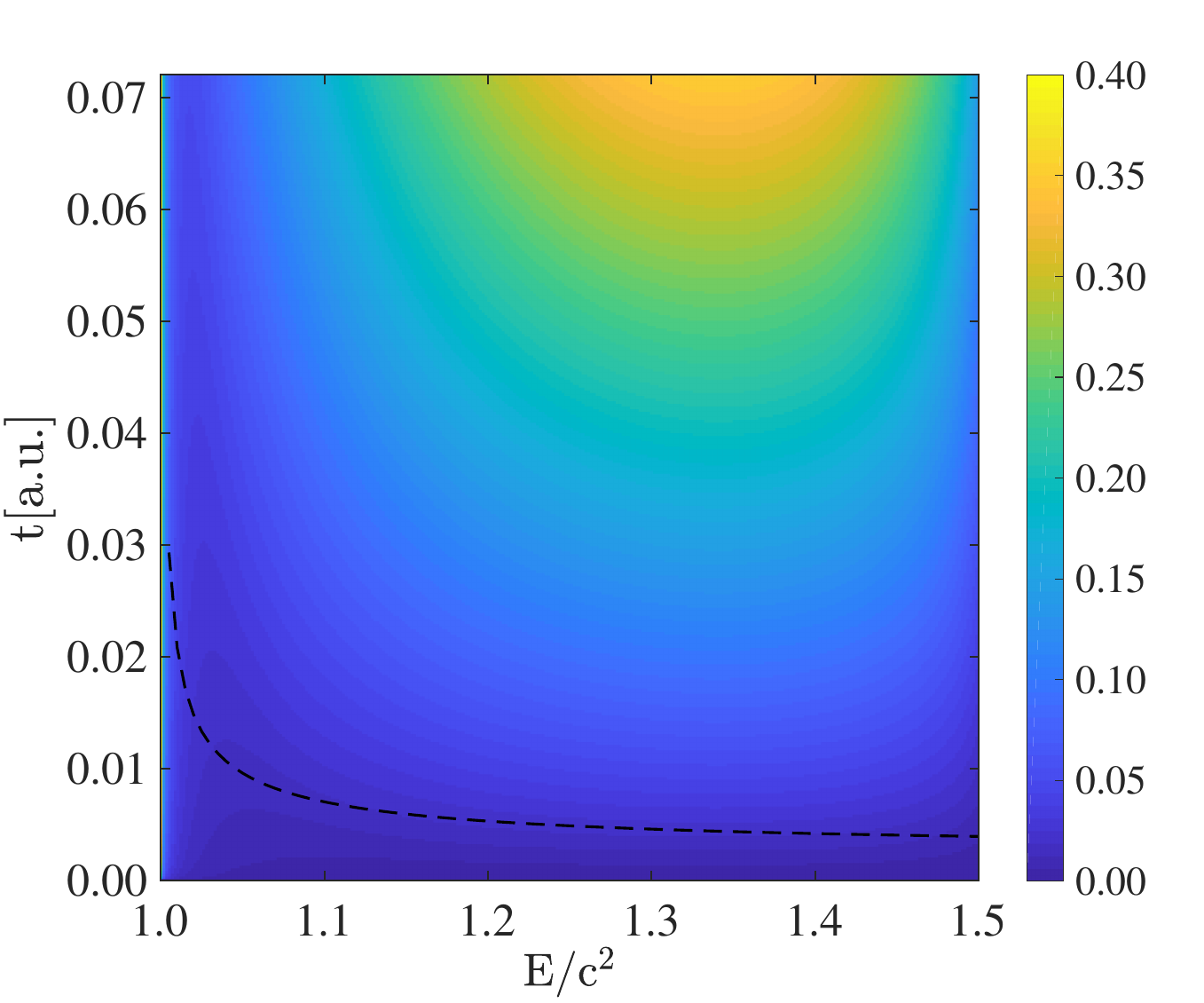}
   	\caption {The energy spectrum of the created electrons for the most probable longitudinal momentum $\rm p_\parallel=0$ as a function of time for Case I. The other parameters are the same as in Fig.~\ref{fig:Figure1}. } 	
	\label{fig:fig5-caseI}
\end{figure}

\subsection{Quantum transmission coefficient for case III } 
	\label{qtc3}

In comparison with case II, this section will discuss the field combination where the vector potential $\rm A_y(x)$ is located at $\rm x_B=-L$ and the scalar potential $\rm \phi(x)$ is still located at $\rm x=0$. In this combination, the right-propagating wave function will first be scattered by the vector potential $\rm A_y(x)$ and then by the scalar potential $\rm \phi(x)$. After two scatterings, the wave function is also divided into three parts by the two potentials at different regions as follows
\begin{equation} 
	\label{eq:psi_iii}
	\rm \psi(x)=
	\begin{cases}
		\rm \psi_{+,p_{i,\perp},+1/2}(x)+ r \psi_{+,-p_{i,\perp},+1/2}(x), & \rm  x<-L \\
		\rm c_1 \psi_{+,p_{2,\perp},+1/2}(x)+ c_2 \psi_{+,-p_{2,\perp},+1/2}(x), & \rm -L<x<0 \\
		\rm  t \psi_{-,p_{f,\perp},+1/2}(x), & \rm x>0.
	\end{cases}
\end{equation}
The wave function is continuous at two places $\rm x=-L$ and $\rm x=0$. Similarly to case II above, we can obtain the following system of equations
\begin{eqnarray}
	\label{eq:con_eq2}
	\begin{aligned} 
        &  \rm \sqrt{\rm \frac{E_2 (E_i + c^2)}{E_i (E_2 + c^2)}} \big[e^{-i p_{i,\perp} L} + r \, e^{i p_{i,\perp} L} \big]=c_1 e^{-i p_{2,\perp} L} + c_2 e^{i p_{2,\perp} L} \,, \\
        & \rm \frac{c_1 + c_2}{c} = t \sqrt{\rm \frac{E_2 (E_f + c^2)}{E_f (E_2 + c^2)}} \frac{ p_{f,\perp} - i p_{f,\parallel}}{E_f + c^2} \,, \\ 
        & \rm \sqrt{\rm \frac{E_2 (E_i + c^2)}{E_i (E_2 + c^2)}} \big[\frac{p_{i,\perp} + i p_{i,\parallel}}{E_i + c^2} e^{-i p_{i,\perp} L} - r \frac{p_{i,\perp} - i p_{i,\parallel}}{E_i + c^2} e^{i p_{i,\perp} L} \big] \\
		& \rm =c_1 e^{-i p_{2,\perp} L} \frac{p_{2,\perp} + i p_{2,\parallel}}{E_2 + c^2} - c_2 e^{i p_{2,\perp} L} \frac{ p_{2,\perp} - i p_{2,\parallel}}{E_2 + c^2} \,, \\
        & \rm c_1 \frac{c (p_{2,\perp} + i p_{2,\parallel})}{E_2 + c^2} - c_2 \frac{ c (p_{2,\perp} - i p_{2,\parallel})}{E_2 + c^2} = t \sqrt{\rm \frac{E_2 (E_f + m c^2)}{E_f (E_2 + m c^2)}},
	\end{aligned} 
\end{eqnarray}
where the momentum and energy in different zones are like $\rm cp_{i,\perp}=(E^2-c^2p_{i,\parallel}^2-c^4)^{1/2}$ for region $\rm R_1$, $\rm E_2=E_i$, $\rm c p_{2,\perp}=(E_2^2-c^2p_{2,\parallel}^2-c^4)^{1/2}$ for region $\rm R_2$ with $\rm p_{2,\parallel}=p_{i,\parallel}-eA_0$, and $\rm E_f=e\phi_0-E_2$, $\rm c p_{f,\perp}=(E_f^2-c^2p_{f,\parallel}^2-c^4)^{1/2}$ for region $\rm R_3$ with $\rm p_{f,\parallel}=p_{2,\parallel}$. Solving this system of equations can give the transmission coefficient
\begin{eqnarray}
	\label{eq:tran_coef3}
	\begin{aligned}
		\rm T_{III}(E_i) & =  \rm [4c^2 p_{i,\perp} p_{2,\perp}^2 p_{f,\perp} (E_i+c^2)(E_f+c^2)]/  \\ 
		&\{\rm [c_a' \sin(\eta)+c_b' \cos(\eta)]^2+[c_c' \sin(\eta)-c_d' \cos(\eta)]^2\},
	\end{aligned}
\end{eqnarray}
with the coefficients as $\rm c_a' = p_{i,\perp}(E_f+c^2)(E_2+c^2)+p_{f,\perp}(E_i+c^2)(E_2-c^2)$, 
$\rm c_b' = p_{2,\perp} c^2 (p_{i,\parallel} p_{f,\perp}+p_{i,\perp} p_{f,\parallel})$, $\rm c_c' =p_{i,\parallel} (E_f+c^2)(E_2+c^2)-p_{2,\parallel}(E_i+c^2)(E_f+c^2)+p_{f,\parallel}(E_i+c^2)(E_2-c^2)$, and $ \rm c_d' = p_{2,\perp}[(E_i+c^2)(E_f+c^2)+c^2 (p_{i,\perp} p_{f,\perp}-p_{i,\parallel} p_{f,\parallel})]$.

\section*{Appendix II: The energy spectra of the created electrons for Case I and III}
\begin{figure}%[htbp]
	\centering
	\includegraphics[height=7.5cm,width=8.5cm]{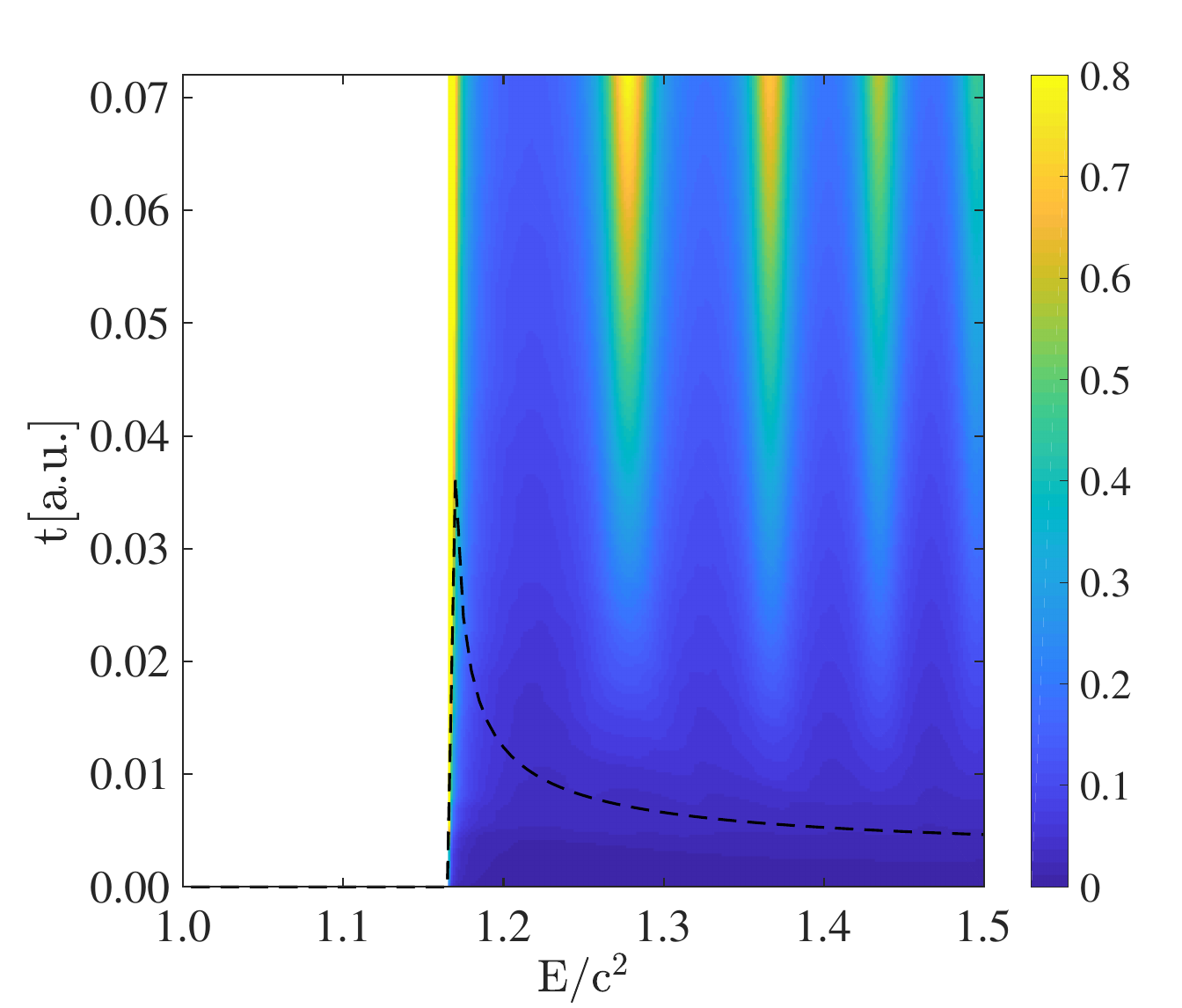}
   	\caption {The energy spectrum of the created electrons for the most probable longitudinal momentum $\rm p_\parallel=eA_0/c=0.6c$ as a function of time for Case III. The other parameters are the same as in Fig.~\ref{fig:Figure1}. } 	
	\label{fig:fig5-caseIII}
\end{figure}

In this section, we present the time-resolved spectra of the produced electrons, serving as a supplement to the energy spectra shown in Fig.~5 of the main text. These results correspond to Case I and Case III, respectively, and provide further insight into the spatiotemporal dynamics of pair production. In each figure, the black dashed curve indicates the time when the corresponding positrons turn back at the vector potential ($\rm x=L$) and arrive at $\rm x=0$. The other simulation parameters are identical to those used in Fig.~1.

\begin{figure}%[htbp]
	\centering
	\includegraphics[height=4.5cm,width=8.5cm]{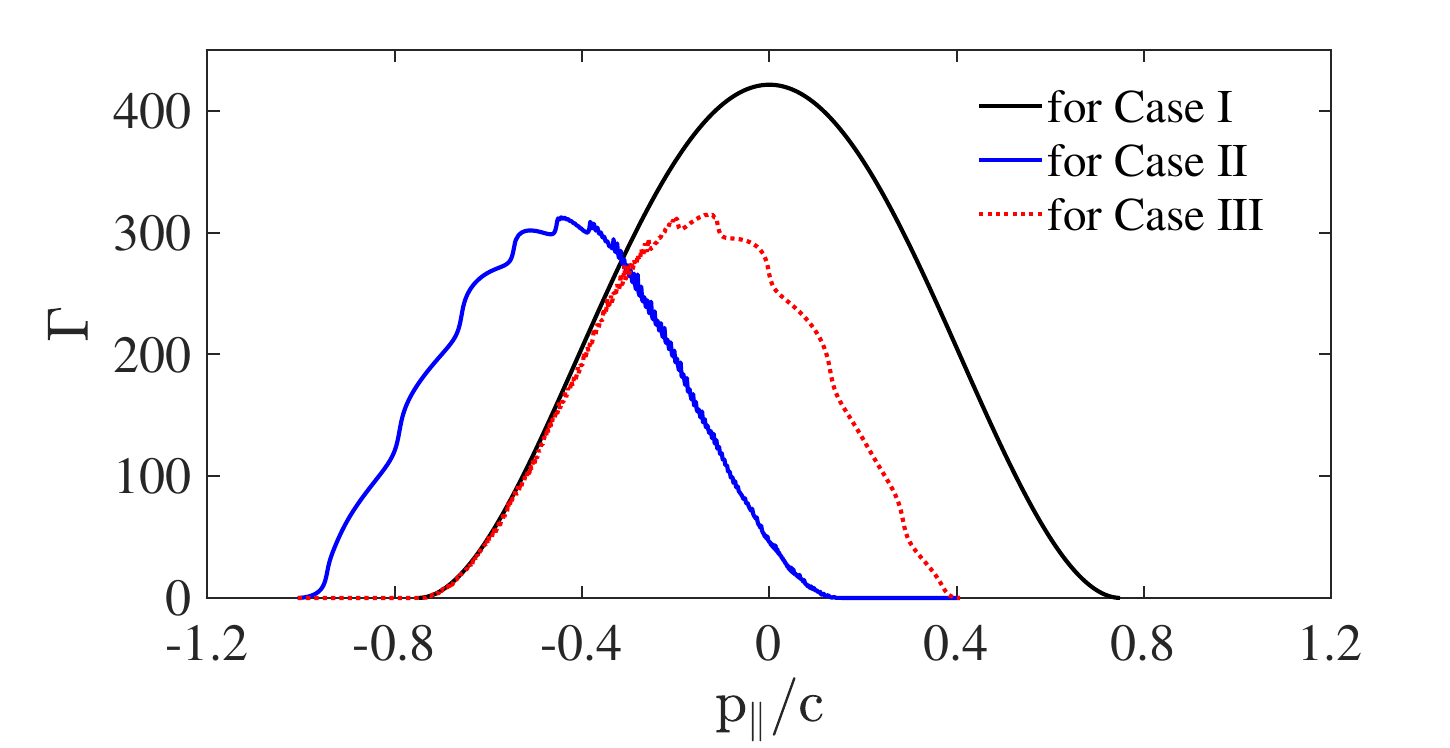}
   	\caption {The creation rate $\Gamma$ as a function of momentum $\rm p_\parallel$. The other parameters are the same as in Fig.~\ref{fig:Figure1}. } 	
	\label{fig:py-gamma}
\end{figure}

The two heatmaps clearly reveal that the Klein region shows no temporal dependence throughout the interaction. This observation suggests that the Klein region is not a result of particle reflection or dynamical trapping induced by the external vector potential, but rather a direct consequence of the instantaneous change of the phase structure of virtual particles by the electromagnetic vector potential.

For Case I [see Fig.~\ref{fig:fig5-caseI}], the Klein region extends from $\rm c^2$ to $\rm 1.5c^2$, in agreement with Fig.~3(d) of the main text. For Case III [see Fig.~\ref{fig:fig5-caseIII}], the energy of the created particles ranges from $\rm 1.33c^2$ to $\rm 1.5c^2$, consistent with Fig.~3(f) in the main text. The remarkable consistency of these energy regions across both cases and their temporal invariance provide compelling evidence that the Klein region is a static feature determined by the field configuration, rather than a dynamical outcome of particle trajectories.

\section*{Appendix III: The pair creation rate for different $p_\parallel$}

This section presents the momentum-resolved pair production rates, as shown in Fig.~\ref{fig:py-gamma}. The black, blue, and red curves correspond to Case I, Case II, and Case III as in Fig.~1 of the main text, respectively. By integrating the momentum-dependent production rates over all $\rm p_\parallel$, one obtains the total creation rate as plotted in Fig.~2 of the main text. Among the three cases, Case I yields the highest total production rate, reaching 7251. The total yields for Case II and Case III are similar, with values of 4260 and 4244, respectively. However, it is worth noting that despite the nearly identical total pair yields in Case II and Case III, Fig.~\ref{fig:py-gamma} clearly shows that the optimal momentum $\rm p_\parallel$ at which pair production is marked different between the two cases.

\bibliography{ab-effect_ref.bib}

\end{document}